\documentclass[english,keywords,amsmath,amssymb,twocolumn]{revtex4}
\usepackage[T1]{fontenc}
\usepackage[latin1]{inputenc}
\usepackage{braket}
\usepackage{babel}%
\usepackage{graphicx}
\usepackage{color}
\usepackage{bm}
\usepackage{longtable}
\usepackage{amsmath}
\usepackage{amsfonts}
\usepackage{dsfont}
\usepackage{amssymb}
\usepackage{hyperref}
\begin{document}

\title{All entangled states are quantum coherent with locally distinguishable
bases}
\author{Asmitha Mekala and Ujjwal Sen}
\affiliation{Harish-Chandra Research Institute, HBNI, Chhatnag Road, Allahabad 211 019, India}
\begin{abstract}
We find that a bipartite quantum state is entangled if and only if it is quantum coherent with respect to complete bases of states in the corresponding system that are distinguishable under local quantum operations and classical communication. The corresponding minimal quantum coherence is the entanglement of formation.  Connections to the relative entropy of entanglement and quantum coherence, and generalizations to the multiparty case are also considered.
\end{abstract}
\maketitle

\section{Introduction}
\label{bilambita-loi}

Entanglement \cite{lichu} and coherence \cite{ja-re-ja-re-uRe} of quantum states result from the superposition principle, except that for the former, at least a two-party situation is necessary. Both have been used to develop resource theories. 
Indeed, entanglement and quantum coherence are the main
resources in many applications of quantum information for demonstrating better
performances than their classical counterparts.
It is also known that an entangling gate necessarily requires coherence at the input to produce entangled states. 

The literature of local distinguishability or its absence of sets of orthogonal states of multiparty systems have developed rather independently. Indeed, it was noticed that entanglement of the constituent states is not directly related to the local indistinguishability of such sets - at least, not in all cases \cite{sakal-shunya-kare-ek, sakal-shunya-kare-dui, knaThaler-aTha}.

The problem of whether or not a state is entangled is known to be intricate and has as yet not been solved. The situation is similar for quantum coherence with respect to clusters of bases with specific properties, and for local distinguishability of sets of quantum states. The quantifications of these concepts do exist in the literature \cite{lichu, ja-re-ja-re-uRe, dukher-rajani-prabhat-hoi-na}, but are typically difficult to compute.

Here we show that entangled quantum states can be seen as quantum coherent states in locally distinguishable bases. Moreover, a convex-roof based measure of quantum coherence, of a bipartite quantum state of arbitrary dimensions, in optimal locally distinguishable bases turn out to be the entanglement of formation of the state, where the latter is a measure of entanglement \cite{boRda}. A different approach of quantification of quantum coherence, for a bipartite state, using the concept of relative entropy \cite{ke-je-bale-debe}, provides an upper bound for relative entropy of entanglement of the state, where the latter is another measure of entanglement \cite{golam-chor, golam-chor-1}. We then show that the considerations can be carried over to multiparty systems. We therefore find that the two salient resource theories of quantum information, viz. entanglement and quantum coherence, are closely related, and that the relation is effected by using  \emph{a priori} unrelated concepts in the domain of local distinguishability of sets of orthogonal multiparty states.

\section{Definitions and results}
\label{abar-asiba-phire}

We will require the concepts of von Neumann entropy and relative entropy between quantum states \cite{ke-je-bale-debe}. 
The von Neumann entropy of a quantum state \(\varrho\) is denoted by \(S(\varrho)\) and is given by 
\begin{equation}
S(\varrho) = -\mbox{tr}(\varrho \log_2 \varrho).
\end{equation}
The von Neumann relative entropy between two quantum states, \(\varrho\) and \(\varsigma\), is denoted by \(S(\varrho \parallel \varsigma) \), and is given by 
\begin{equation}
S(\varrho \parallel \varsigma) = \mbox{tr}(\varrho \log_2 \varrho - \varrho \log_2 \varsigma).
\end{equation}
It is to be noted that the relative entropy is not symmetric with respect to its arguments.

A qualitative definition of quantum coherence, as  has already been given in the literature, can be as follows \cite{ja-re-ja-re-uRe}.\\
\noindent \textbf{Definition.} A pure quantum state \(|\psi\rangle\) of a physical system represented by a Hilbert space \(\mathbb{C}^d\) is said to be quantum coherent with respect to a complete orthonormal basis of \(\mathbb{C}^d\) if it is not an element of that basis.

The notion has also been quantified, and one of the quantifications is as follows \cite{ja-re-ja-re-uRe}.\\
\noindent \textbf{Definition.} Let \(B\) be a 
complete orthonormal basis of pure states in \(\mathbb{C}^{d}\).
Let \(C_B(|\psi\rangle)\) be the relative entropy of quantum coherence of \(|\psi\rangle \in \mathbb{C}^{d}\),
so that 
\begin{equation}
C_B(|\psi\rangle) = \min_{\rho_B \in M_B} S(|\psi\rangle \langle \psi| \parallel \rho_B),
\end{equation}
where \(M_B\) is the set of all probabilistic mixtures of the projectors onto the elements of \(B\). 

Complete orthonormal bases of bipartite quantum systems are of course distinguishable under global operations. One just makes a measurement onto that basis. Things are more complicated however when a restricted class of operations is allowed. An important such restricted class is the class of local quantum operations and classical communication (LOCC) \cite{boRda,golam-chor-1,neel-beRechhe}. If a complete orthonormal basis is 
also distinguishable under LOCC, we will call the basis as ``locally distinguishable''.

A bipartite pure state is said to be entangled if it cannot be written as a tensor product of pure states of the two systems.\\
\noindent \textbf{Theorem 1.} \emph{A bipartite pure state is entangled if and only if it has a nonzero quantum coherence with respect to all locally distinguishable complete orthonormal bases.}

\noindent \texttt{Proof.}
Let \(|\psi\rangle\) be a pure entangled state of a bipartite quantum system, the Hilbert space corresponding to which is \(\mathbb{C}^{d_1} \otimes \mathbb{C}^{d_2}\). 
Let us now consider an arbitrary locally distinguishable complete orthonormal basis. 
\(|\psi\rangle\) will have vanishing quantum coherence in this basis if and only if it is an element of this basis. 
However, if \(|\psi\rangle\) is an element of this basis, the latter cannot be locally distinguishable, as it was proven in Ref. 
\cite{knaThaler-aTha} that any complete orthonormal basis containing even a single entangled state cannot be locally distinguishable.
Therefore, \(|\psi\rangle\) must have a nonzero quantum coherence in any locally distinguishable complete orthonormal  basis.
%
%

On the other hand, an arbitrary pure product state can always be expanded to form a complete bi-orthonormal product basis, which can always be distinguished by LOCC.
Therefore, the product state has zero quantum coherence at least with respect to this basis. \hfill{}\(\blacksquare\)

It is clear that the states formed by mixing locally distinguishable orthonormal complete basis states are separable states, and the basis forms the spectral states of the separable state, where a separable state is any state that can be prepared by LOCC after starting from product states \cite{megh}. Not all separable states are however of this type, 
that is, not all separable states have a spectral basis that is locally distinguishable. To see this, let us consider a specific example. Consider the two-qubit states, \(|\psi^\pm\rangle = (|01\rangle \pm |10\rangle)/\sqrt{2}\), and an unequal mixture of them, viz. \(\rho_2=p|\psi^+\rangle \langle \psi^+| + (1-p) |\psi^-\rangle \langle \psi^-|\), where \(p\in (0,1)\) and \(p \ne 1/2\). By using the positive partial transpose (PPT) criterion \cite{sreeradha}, one can check that this family has only entangled states. (The state for \(p=1/2\) is separable.)
Now, the spectral decomposition of \(\rho_2\) is unique, as there is no degeneracy in the spectrum. Also, a locally distinguishable basis is necessarily an orthogonal set. The spectral basis of \(\rho_2\) is incomplete, but can always be completed to a full orthonormal basis, and any such completion - for \(p \ne 1/2\) - will be locally indistinguishable, as at least two states of it will be entangled, viz. \(|\psi^\pm\rangle\) \cite{knaThaler-aTha}. Therefore, there is no locally distinguishable basis which can be probabilistically mixed to form \(\rho_2\) for any 
\(p \ne 1/2\). 
Because of the existence of such examples, it may seem that quantum coherence of a bipartite state with respect to locally distinguishable bases may not be related to the state's entanglement content. We however have the following result.\\
%
\noindent \textbf{Theorem 2.} \emph{The minimum among quantum coherences with respect to all locally distinguishable complete orthonormal bases of any bipartite pure quantum state is given by its local von Neumann entropy.}

\noindent \texttt{Proof.} The minimal relative entropy distance of a pure state \(|\psi\rangle\) of \(\mathbb{C}^{d_1} \otimes \mathbb{C}^{d_2}\), written in Schmidt decomposition as \(\sum_{i=1}^n\alpha_i|ii\rangle\), from the set of separable states, is attained in the state \(\sum_{i=1}^n\alpha_i^2 |ii\rangle \langle ii|\), and is given by the von Neumann entropy of either of the local densities \cite{golam-chor, golam-chor-1}. Here, \(n \leq \min\{d_1, d_2\}\), and \(\alpha_i\) are real and positive. The state \(\sum_{i=1}^n\alpha_i^2 |ii\rangle \langle ii|\) is of course a separable state, but is also a mixture of states of a locally distinguishable orthonormal basis, viz. \(\{|i\rangle|j\rangle\}_{i=1, j=1}^{d_1, d_2}\). \hfill{} \(\blacksquare\)

The result reminds us of similar ones for quantum discord \cite{peyara} and quantum work deficit \cite{narkol}, which also were equal to the local von Neumann entropy for all pure bipartite states.

Natural extensions of the concept of quantum coherence to mixed states can be made in several ways. One of them is by using the concept of the convex roof \cite{boRda}. 
\\
\noindent \textbf{Definition.} A quantum state \(\rho\) on \(\mathbb{C}^d\) is said to be quantum coherent with respect to a class of complete orthonormal bases \(\{B\}\) of \(\mathbb{C}^d\) with a special pre-settled property if it cannot be written as a convex (i.e., probabilistic) sum of pure states of \(\mathbb{C}^d\) with zero minimal quantum coherence when optimized over such bases.\\
The qualitative definition of quantum coherence can be quantified as follows.\\
\noindent \textbf{Definition.} For a quantum state \(\rho\) on \(\mathbb{C}^{d}\),
its quantum coherence with respect to a class \(\{B\}\) of bases on \(\mathbb{C}^{d}\) is given by
\begin{equation}
C_{\{B\}}(\rho) =  \min \sum_i p_i \min_{B\in\{B\}}C_B (|\psi_i\rangle), 
\end{equation}
where the outer minimization is over all decompositions of \(\rho\) into \(\sum_i p_i |\psi_i\rangle \langle \psi_i|\).\\
When the set \(\{B\}\) contains all bases of \(\mathbb{C}^d\), it is clear that all quantum states will have vanishing quantum coherence. Quantum coherence is typically defined with respect to a fixed basis, and one subsequently demonstrates that it satisfies certain conditions \cite{ja-re-ja-re-uRe}. The definition presented above, however, involves a class of bases. As we show in the Supplementary Material, it can also be shown to satisfy the usual conditions of a quantum coherence measure, in the case of interest to us, viz. when \(\{B\}\) is the class of all locally distinguishable complete orthonormal bases of a multiparty quantum system.

We will now need the concept of entanglement of formation \cite{boRda},
which, for
a bipartite quantum state \(\rho_{AB}\) is defined as 
\begin{equation}
\label{rajakini-rami}
E_F(\rho_{AB}) = \min \sum_i p_i E_F (|\psi_i\rangle_{AB}),
\end{equation}
where the minimization is over all decompositions of \(\rho_{AB}\) into \(\sum_i p_i |\psi_i\rangle \langle \psi_i|\), and where the entanglement of formation for a pure bipartite state is given by the von Neumann entropy of either of the local densities \cite{rajani}.
   Entanglement of formation has been put forward as a measure of entanglement and is typically difficult to compute \cite{boRda, damodar}, where a bipartite entangled state is one which is not separable.

At the qualitative level, we have the following result, the proof of which is presented in the Supplementary Material.\\ 
\noindent \textbf{Theorem 3.} \emph{A bipartite quantum state, possibly mixed, is entangled if and only if it has a nonzero quantum coherence with respect to all locally distinguishable complete orthonormal bases.}


Just like for pure states, the connection between entanglement and quantum coherence in LOCC-distinguishable bases can be taken to a quantitative level.\\
\noindent \textbf{Theorem 4.} \emph{The quantum coherence in locally distinguishable bases of a bipartite quantum state, possibly mixed, is the entanglement of formation of the state.}\\
A proof is presented in the Supplementary Material. A similar result was obtained in Ref. \cite{asbo-arek-din}, where quantum coherence in product bases and its convex-roof extension were considered. 
We note that a complete orthonormal basis having even a single entangled state is necessarily locally indistinguishable \cite{knaThaler-aTha}. 
However, there exists complete orthonormal bases of product states that are locally indistinguishable  \cite{sakal-shunya-kare-ek}. 
We also remember here that an entangling gate necessarily requires coherence at the input to produce entangled states at the output. 
See \cite{shono-kono-ek-din} in this regard. 


As already alluded to, there are other avenues of natural extensions of the concept of quantum coherence to mixed quantum states. One such is given as follows \cite{ja-re-ja-re-uRe}.\\
\noindent \textbf{Definition.} A quantum state \(\rho\) on \(\mathbb{C}^d\) is said to be relative quantum coherent with respect to a complete orthonormal basis \(B\) of \(\mathbb{C}^d\) if it is not 
a mixture of states of the basis.\\
The christening is non-standard, and is made to distinguish it from the previous definition, in this section, of quantum coherence,
and is chosen because it is relative to a particular basis \(B\) and not, as in the previous case, with respect to a class \(\{B\}\) of bases.
%
%
We now provide a quantification of the notion of relative quantum coherence \cite{ja-re-ja-re-uRe}.\\
\noindent \textbf{Definition.} For a quantum state \(\rho\) on \(\mathbb{C}^{d}\),
its relative entropy of quantum coherence with respect to the basis \(B\) on \(\mathbb{C}^{d}\) is given by
\begin{equation}
C_B^R(\rho) = \min_{\rho_B \in M_B} S(\rho \parallel \rho_B).
\end{equation}
This definition of quantum coherence has been widely used to build a resource theory, and the corresponding  monotonicity properties have been proven in the literature \cite{ja-re-ja-re-uRe}.


We will now need the concept of 
the relative entropy of entanglement \cite{golam-chor, golam-chor-1}, which, for
%
a bipartite state \(\rho_{AB}\) is given by the minimal relative entropy distance of the state from the set of separable states in the same Hilbert space, so that 
\begin{equation}
E_R(\rho_{AB}) = \min_{\sigma_{AB}} S(\rho_{AB} \parallel \sigma_{AB}), 
\end{equation}
where \(\sigma_{AB}\) is a separable state.
Just like the entanglement of formation, the relative entropy of entanglement has also been proposed as a 
measure of entanglement, and is again typically difficult to compute \cite{golam-chor, golam-chor-1}.

At the qualitative level, we have the following result.\\ 
\noindent \textbf{Theorem 5.} \emph{Any bipartite entangled state, possibly mixed, has a nonzero relative quantum coherence with respect to all locally distinguishable complete orthonormal bases.}


And on the quantitative level, we have the following relation.\\
\noindent \textbf{Theorem 6.} \emph{The minimal relative entropy of quantum coherence of a bipartite quantum state, possibly mixed, with locally distinguishable bases is bounded below by the relative entropy of entanglement of the state.}\\
The proofs of Theorems 5 and 6 are presented in the Supplementary Material.



We have until now been considering the case of entanglement of bipartite states and local distinguishability of sets of bipartite states. These considerations can be carried over to the multiparty case. 
Both the concepts, viz. entanglement and local distinguishability, are far richer in the  multiparty domain. 
The connection between entanglement and quantum coherence can however be carried over to the multiparty case, and we exemplify the situation by considering two diametrically opposite types of multiparty entanglements in the following two theorems (proofs in Supplementary Material).\\
\noindent \textbf{Theorem 7.} \emph{A multiparty pure state in}
\(\mathbb{C}^{d_1} \otimes \mathbb{C}^{d_2} \otimes \ldots \mathbb{C}^{d_m}\) \emph{is entangled across at least one bi-partition if and only if it is quantum coherent with respect to all locally distinguishable complete orthonormal bases.} \\
\noindent \textbf{Theorem 8.} \emph{A multiparty pure state in} \(\mathbb{C}^{d_1} \otimes \mathbb{C}^{d_2} \otimes \ldots \otimes \mathbb{C}^{d_m}\) \emph{is genuinely multiparty entangled if and only if it is quantum coherent with respect to complete orthonormal bases that are locally distinguishable in at least one bipartition of the} \(m\) \emph{parties.}\\
These results in the multiparty scenario are for pure states, but the generalizations to the regime of mixed states are similar to those already done in the bipartite case.


It is important to mention here about the experimental feasibility of measuring the quantities examined. Entanglement and quantum coherence are deeply studied topics and their experimental characterization and quantification have been performed in the literature in a large number of works. These strategies are probably more studied for entanglement than for quantum coherence. In particular, the quantum coherence measures studied here have not been experimentally characterized in the literature. 
However, these quantum coherence measures have all been found to be intimately related, in many cases equal, to known measures of entanglement.  And the latter have been characterized and quantified in the literature.

\section{Conclusion}
%
\label{garhita-kaj}

Entanglement of shared quantum states forms one of the most successful resources for performing quantum information tasks \cite{lichu}. It is therefore interesting to characterize and quantify it in as many different ways as possible, as that may lead to a deeper understanding of the concept and also potentially result in new applications. 

Quantum coherence has also been argued to be resourceful in attaining quantum advantage in specified tasks over their classical counterparts \cite{ja-re-ja-re-uRe}. 

We found that quantum coherence in locally distinguishable bases can be used to define and quantify entanglement. It is to be noted that there exists complete globally distinguishable bases that are not locally distinguishable, and it has typically been argued that the local indistinguishability of such bases is unrelated to the entanglement content of the constituent states \cite{sakal-shunya-kare-ek,sakal-shunya-kare-dui,knaThaler-aTha}. 

We initially proved the results for pure bipartite states. Depending the way, the generalization of quantum coherence to mixed states is executed, the relation between quantum coherence in locally distinguishable bases and entanglement is obtained in terms of the entanglement of formation \cite{boRda} or the relative entropy of entanglement \cite{golam-chor, golam-chor-1}. The results were then shown to be generalizable to the multiparty domain.



\begin{center}
\textbf{Supplementary Material for\\
All entangled states are quantum coherent with locally distinguishable
bases}

\texttt{Asmitha Mekala and Ujjwal Sen}\\
\texttt{Harish-Chandra Research Institute, HBNI, Chhatnag Road, Allahabad 211 019, India}

\end{center}


\section{Quantum coherence}

Let us first repeat a few definitions that are already present in the main text \cite{ja-re-ja-re-uRe-1}.



\textbf{Definition.} A quantum state \(\rho\) on \(\mathbb{C}^d\) is said to be quantum coherent with respect to a class of complete orthonormal bases \(\{B\}\) of \(\mathbb{C}^d\) with a special pre-settled property if it cannot be written as a convex (i.e., probabilistic) sum of pure states of \(\mathbb{C}^d\) with zero minimal quantum coherence when optimized over such bases.

The qualitative definition of quantum coherence can be quantified as follows.\\
\noindent \textbf{Definition.} For a quantum state \(\rho\) on \(\mathbb{C}^{d}\),
its quantum coherence with respect to a class \(\{B\}\) of bases on \(\mathbb{C}^{d}\) is given by
\begin{equation}
C_{\{B\}}(\rho) =  \min \sum_i p_i \min_{B\in\{B\}}C_B (|\psi_i\rangle), 
\end{equation}
where the outer minimization is over all decompositions of \(\rho\) into \(\sum_i p_i |\psi_i\rangle \langle \psi_i|\).

When the set \(\{B\}\) contains all bases of \(\mathbb{C}^d\), it is clear that all quantum states will have vanishing quantum coherence. 

A resource theory of quantum coherence requires the identification of free operations, free states, and a measure that satisfies monotonicity properties with respect to the free operations and vanishes for free states. Our interest in this manuscript lies in multiparty quantum systems. Let us, for definiteness, consider bipartite 
quantum states, although many of the considerations below hold equally well for other multiparty systems. Also, we wish to consider the class \(\{B\}\) as the class of all locally distinguishable complete orthonormal bases. Let us denote this class as \(\{B_L\}\).

We prove in Theorem 4 that \(C_{\{B_L\}}(\rho)\) for a bipartite quantum state \(\rho\) is equal to its entanglement of formation, \(E_F(\rho)\). Now \(E_F(\rho) = 0\) holds if and only if \(\rho\) is separable \cite{boRda-1}. Therefore, we identify the separable states as the set of free states in our resource theory, for which the measure is \(C_{\{B_L\}}\). We still need to identify the free operations, which maps free states to free states, and with respect to which the measure \(C_{\{B_L\}}(\rho)\) will be a monotone. 
A natural class of operations to consider for a bipartite quantum system and which is also operationally important, is the LOCC (local quantum operations and classical communication) class. 
Entanglement of formation, and hence our measure of quantum coherence, is monotonically non-increasing, on average, for this class \cite{boRda-1}. 
The general class of free operations in our resource theory is however bigger, and includes separable superoperators \cite{neelmadhab}, and also the swap operator (of the entire subsystems). 
These latter ones are however not implementable by using LOCC, the natural class of operations in the distant laboratories paradigm.


\section{Proof of Theorem 3}

\noindent\textbf{Theorem 3.} \emph{A bipartite quantum state, possibly mixed, is entangled if and only if it has a nonzero quantum coherence with respect to all locally distinguishable complete orthonormal bases.}

\noindent\texttt{Proof.} Let \(\rho\) be a bipartite entangled state on a physical system represented by the Hilbert space \(\mathbb{C}^{d_1} \otimes \mathbb{C}^{d_2}\). 
Since this state, which is
possibly mixed, is entangled,  every probabilistic decomposition of it into pure states contains at least one entangled (pure) state \cite{megh-1}. Let \(\{B_L\}\) be the set of all locally distinguishable complete orthonormal bases on \(\mathbb{C}^{d_1} \otimes \mathbb{C}^{d_2}\). 
We have proved in Theorem 1 that an entangled pure state is quantum coherent with respect to all bases in \(\{B_L\}\). Therefore, every decomposition of \(\rho\) into pure states will contain at least one (pure) state that is quantum coherent with respect to all bases in \(\{B_L\}\). 

Let us now assume that \(\rho\) is quantum coherent with respect to all bases in \(\{B_L\}\). Therefore, by (the convex-roof) definition of quantum coherence, it follows that it cannot be written as a convex sum of pure states, all of which have vanishing quantum coherence when optimized over bases in \(\{B_L\}\). Therefore, by Theorem 1, any convex decomposition of \(\rho\) will contain a pure entangled state, implying that \(\rho\) is entangled.
\hfill{} \(\blacksquare\)

\section{Proof of Theorem 4}

\noindent \textbf{Theorem 4.} \emph{The quantum coherence in locally distinguishable bases of a bipartite quantum state, possibly mixed, is the entanglement of formation of the state.}

\noindent\texttt{Proof.}
Let \(\rho\) be a bipartite entangled state on a physical system represented by the Hilbert space \(\mathbb{C}^{d_1} \otimes \mathbb{C}^{d_2}\). 
Let \(\{B_L\}\) be the set of all locally distinguishable complete orthonormal bases on \(\mathbb{C}^{d_1} \otimes \mathbb{C}^{d_2}\).
Let \(\rho = \sum_i p_i |\psi_i\rangle \langle \psi_i|\) be decomposition of \(\rho\), where \(|\psi_i\rangle\)
are pure states of \(\mathbb{C}^{d_1} \otimes \mathbb{C}^{d_2}\), and where \(p_i\) forms a probability distribution.
Now, 
\begin{equation}
C_{\{B_L\}}(\rho) =  \min \sum_i p_i \min_{B_L\in\{B_L\}}C_{B_L} (|\psi_i\rangle), 
\end{equation}
where the outer minimization is over all convex decompositions of \(\rho\) into pure states. Using Theorem 2, we can write 
\(\min_{B_L\in\{B_L\}}C_{B_L} (|\psi_i\rangle)\) as \(E_F(|\psi_i\rangle)\), so that we have 
\begin{equation}
C_{\{B_L\}}(\rho) =  \min \sum_i p_i E_F(|\psi_i\rangle), 
\end{equation}
with the right-hand-side being exactly equal to the entanglement of formation of \(\rho\). 
\hfill{} \(\blacksquare\)

\section{Proofs of Theorems 5 and 6}

\noindent \textbf{Theorem 5.} \emph{Any bipartite entangled state, possibly mixed, has a nonzero relative quantum coherence with respect to all locally distinguishable complete orthonormal bases.}

\noindent \texttt{Proof.} 
Let \(\rho\) be a bipartite entangled state on a physical system represented by the Hilbert space \(\mathbb{C}^{d_1} \otimes \mathbb{C}^{d_2}\). 
Since this state, which is
possibly mixed, is entangled,  every probabilistic decomposition of it into pure states contains at least one entangled (pure) state \cite{megh-1}. Let \(\{B_L\}\) be the set of all locally distinguishable complete orthonormal bases on \(\mathbb{C}^{d_1} \otimes \mathbb{C}^{d_2}\). 
Any probabilistic mixture of elements of any \(B_L \in \{B_L\}\) will provide a separable state, as we have noted in the main text, and therefore cannot produce \(\rho\), which has been taken to be entangled.
Therefore, \(\rho\) is relative quantum coherent with respect to all bases in \(\{B_L\}\).
\hfill{} \(\blacksquare\)

\noindent \textbf{Theorem 6.} \emph{The minimal relative entropy of quantum coherence of a bipartite quantum state, possibly mixed, with locally distinguishable bases is bounded below by the relative entropy of entanglement of the state.}

\noindent \texttt{Proof.} 
Let \(\rho\) be a bipartite entangled state on a physical system represented by the Hilbert space \(\mathbb{C}^{d_1} \otimes \mathbb{C}^{d_2}\). 
Let \(\{B_L\}\) be the set of all locally distinguishable complete orthonormal bases on \(\mathbb{C}^{d_1} \otimes \mathbb{C}^{d_2}\).
Now, the relative entropy of quantum coherence with respect to a basis \(B_L \in \{B_L\}\) is given by
\begin{equation}
C_{B_L}^R(\rho) = \min_{\rho_{B_L} \in M_{B_L}} S(\rho \parallel \rho_{B_L}).
\end{equation}
As mentioned in the main text,  \(M_{B_L}\) contains only separable states, and so, 
\begin{equation}
C_{B_L}^R(\rho) =  \min_{\rho_{B_L} \in M_{B_L}} S(\rho \parallel \rho_{B_L}) \geq \min_{\sigma} S(\rho \parallel \sigma),
\end{equation}
for all \(B_L \in \{B_L\}\),
where \(\sigma\) is an arbitrary separable state on \(\mathbb{C}^{d_1} \otimes \mathbb{C}^{d_2}\).
Therefore, by the definition of relative entropy of entanglement, we have
\begin{equation}
\min_{B_L \in \{B_L\}} C_{B_L}^R(\rho) \geq E_R(\rho).
\end{equation}
Hence, the proof. \hfill{} \(\blacksquare\)

\section{Proof of Theorems 7 and 8}

We have until now been considering the case of entanglement of bipartite states and local distinguishability of sets of bipartite states. These considerations can be carried over to the multiparty case. 
Both the concepts, viz. entanglement and local distinguishability, are far richer in the  multiparty domain. 
The connection between entanglement and quantum coherence can however be carried over to the multiparty case, and we exemplify the situation by considering two diametrically opposite types of multiparty entanglements in the following two theorems.

Let us first define a multiparty pure state in \(\mathbb{C}^{d_1} \otimes \mathbb{C}^{d_2} \otimes \ldots \mathbb{C}^{d_m}\) as entangled if it is entangled across at least one bi-partition. Similarly, we define it to be quantum coherent if it is not an element of an LOCC-distinguishable complete orthonormal basis, where the LOCC in the multiparty scenario is local with respect to all the parties and classical communication is allowed between all the parties.\\
\noindent \textbf{Theorem 7.} \emph{A multiparty pure state in}
\(\mathbb{C}^{d_1} \otimes \mathbb{C}^{d_2} \otimes \ldots \mathbb{C}^{d_m}\) \emph{is entangled if and only if it is quantum coherent with respect to all locally distinguishable complete orthonormal bases.}

\noindent \texttt{Proof.}
Let \(|\tilde{\psi}\rangle\) be a pure entangled state on a physical system represented by the Hilbert space \(\mathbb{C}^{d_1} \otimes \mathbb{C}^{d_2} \otimes \ldots \mathbb{C}^{d_m}\), so that, according to the definition accepted in the main text for entangled multiparty states, the state \(|\tilde{\psi}\rangle\) is entangled across at least one bi-partition. 
Let that bi-partition be \(\mathcal{A}:\mathcal{B}\).
Let \(\{\tilde{B}_L\}\) be the set of all locally distinguishable complete orthonormal bases on \(\mathbb{C}^{d_1} \otimes \mathbb{C}^{d_2}
\otimes \ldots \mathbb{C}^{d_m}\).
Therefore, all elements of \(\{\tilde{B}_L\}\)
will be locally distinguishable in the bi-partition
\(\mathcal{A}:\mathcal{B}\). Consequently, by Theorem 1, \(|\tilde{\psi}\rangle\) is quantum coherent with respect to all bases in \(\{\tilde{B}_L\}\).

On the other hand, if 
\(|\tilde{\psi}\rangle\) is of the form 
\(\otimes_{i=d_1}^{d_m}|\psi_{i}\rangle\), then it can always be expanded into a multi-orthonormal product basis of \(\mathbb{C}^{d_1} \otimes \mathbb{C}^{d_2}
\otimes \ldots \mathbb{C}^{d_m}\), which can be distinguished by LOCC with all the \(m\) parties in separate locations.
\hfill{} \(\blacksquare\)

A convex-roof approach for the extension to mixed multiparty quantum states will then provide the result that a multiparty state, possibly mixed, will be entangled (i.e., not ``fully-separable'') if and only if it is quantum coherent with respect to locally distinguishable complete orthonormal bases.

We now move over to the diametrically opposite scenario (for entangled multiparty pure states). Precisely, we change our definition of entangled multiparty pure states to the exact opposite extreme to what was used in Theorem 7. In the literature, such states are called genuinely multiparty entangled states, and we also call it so here.\\
\noindent \textbf{Theorem 8.} \emph{A multiparty pure state in} \(\mathbb{C}^{d_1} \otimes \mathbb{C}^{d_2} \otimes \ldots \otimes \mathbb{C}^{d_m}\) \emph{is genuinely multiparty entangled if and only if it is quantum coherent with respect to complete orthonormal bases that are locally distinguishable in at least one bipartition of the} \(m\) \emph{parties.}

\noindent \texttt{Proof.} Let \(|\psi\rangle\) be a genuinely multisite entangled state of the \(m\) parties. Therefore,           \(|\psi\rangle\) is entangled across all bipartitions \(A:B\) of the \(m\) parties. Let \(\{B_L\}\) be the set of all complete orthonormal bases are locally distinguishable in at least one bipartition of the m parties. For a given element of that set, \(\{B_L\}\), let that partition be \(A_1:B_1\). Then by Theorem 1,        \(|\psi\rangle\) is quantum coherent with respect to that element of \(\{B_L\}\). 

On the other hand, if \(|\psi\rangle\) is of the form \(|\psi{'}_A\rangle \otimes |\psi{''}_B\rangle\) across some bipartition \(A:B\) of the \(m\) parties, then it can always be completed to a complete orthonormal basis of the Hilbert space of the m parties that is locally distinguishable in \(A:B\). 

This completes the proof. \hfill \(\blacksquare\)

Again, a generalization to the case of mixed states is possible.



\begin{thebibliography}{99}

\bibitem{lichu} 
R. Horodecki, P. Horodecki, M. Horodecki, and K. Horodecki, Rev. Mod. Phys. \textbf{81}, 865 (2009);
O. G{\" u}hne and G. T{\' o}th,
Physics Reports \textbf{474}, 1 (2009);
S. Das, T. Chanda, M. Lewenstein, A. Sanpera, A. Sen(De), and U. Sen, \emph{The separability versus entanglement problem}, in \emph{Quantum Information: From Foundations to Quantum Technology Applications}, second edition, eds. D. Bru{\ss} and G. Leuchs (Wiley, Weinheim, 2019),
arXiv:1701.02187 [quant-ph].




\bibitem{ja-re-ja-re-uRe}
J. {\AA}berg, 
arXiv:quant-ph/0612146 (2006);
%
T. Baumgratz, M. Cramer, and M. B. Plenio, 
Phys. Rev. Lett. \textbf{113}, 140401 (2014);
%
A. Winter and D. Yang, 
Phys. Rev. Lett. \textbf{116}, 120404 (2016);
%
A. Streltsov, G. Adesso, and M. B. Plenio, 
Rev. Mod. Phys. {\bf 89}, 041003 (2017).






\bibitem{sakal-shunya-kare-ek} 
C. H. Bennett, D. P. DiVincenzo, C. A. Fuchs, T. Mor, E. Rains, P. W. Shor, J. A. Smolin, and W. K. Wootters,
Phys. Rev. A \textbf{59}, 1070 (1999).

\bibitem{sakal-shunya-kare-dui} J. Walgate, A. J. Short, L. Hardy, and V. Vedral
Phys. Rev. Lett. \textbf{85}, 4972 (2000).


\bibitem{knaThaler-aTha}
 M. Horodecki, A. Sen(De), U. Sen, and K. Horodecki,
Phys. Rev. Lett. \textbf{90}, 047902 (2003).


\bibitem{dukher-rajani-prabhat-hoi-na}
M. Horodecki, A. Sen(De), and U. Sen,
Phys. Rev. A \textbf{75}, 062329 (2007).

\bibitem{boRda} 
C. H. Bennett, D. P. DiVincenzo, J. A. Smolin, and W. K. Wootters,
Phys. Rev. A \textbf{54}, 3824 (1996).


\bibitem{ke-je-bale-debe} 
A. Wehrl, 
Rev. Mod. Phys. \textbf{50}, 221 (1978).



\bibitem{golam-chor} 
V. Vedral, M. B. Plenio, M. A. Rippin, and P. L. Knight, Phys. Rev. Lett. \textbf{78}, 2275 (1997); 
V. Vedral,
Rev. Mod. Phys. \textbf{74}, 197 (2002).

\bibitem{golam-chor-1} V. Vedral  and M. B. Plenio, Phys. Rev. A 57 1619 (1998).

\bibitem{neel-beRechhe} E. M. Rains, quant-ph/9707002.

\bibitem{megh} 
R. F. Werner,
Phys. Rev. A \textbf{40}, 4277 (1989).

\bibitem{sreeradha} A. Peres,
Phys. Rev. Lett. \textbf{77}, 1413 (1996); M. Horodecki, P. Horodecki, and R. Horodecki, Phys. Lett. A \textbf{221}, 1 (1996).




\bibitem{peyara} 
L. Henderson  and V. Vedral,
J. Phys. A: Math. Gen. \textbf{34}, 6899 (2001);
%
H. Ollivier  and W. H. Zurek,
Phys. Rev.
Lett. 88 017901 (2001);
K. Modi, A. Brodutch, H. Cable, T. Paterek, and V. Vedral, 
Rev. Mod. Phys. \textbf{84}, 1655 (2012); 
A. Bera, T. Das, D. Sadhukhan, S. Singha Roy, A. Sen(De), and U. Sen,
Rep. Prog. Phys. \textbf{81}, 024001 (2018).


\bibitem{narkol} 
 J. Oppenheim, M. Horodecki, P. Horodecki, and R. Horodecki,
Phys. Rev. Lett. \textbf{89}, 180402 (2002);
M. Horodecki, K. Horodecki, P. Horodecki, R. Horodecki, J. Oppenheim, A. Sen(De), and U. Sen,
Phys. Rev. Lett. \textbf{90}, 100402 (2003);
M. Horodecki, P. Horodecki, R. Horodecki, J. Oppenheim, A. Sen(De), and U. Sen, and B. Synak,
Phys. Rev. A \textbf{71}, 062307 (2005).


\bibitem{rajani} 
C. H. Bennett, H. J. Bernstein, S. Popescu, and B. Schumacher,
Phys. Rev. A \textbf{53}, 2046 (1996).


\bibitem{damodar}
S. Hill and W. K. Wootters, 
Phys. Rev. Lett. \textbf{78}, 5022 (1997);
W. K. Wootters,
Phys. Rev. Lett. \textbf{80}, 2245 (1998).




\bibitem{asbo-arek-din}  H. Zhu, Z. Ma, Z. Cao, S.-M. Fei, and V. Vedral,
Phys. Rev. A \textbf{96}, 032316 (2017).


\bibitem{shono-kono-ek-din} 
T. R. de Oliveira and A. O. Caldeira, 
arXiv:quant-ph/0608192; 
A. J. Hudson, R. M. Stevenson, A. J. Bennett, R. J. Young, C. A. Nicoll, P. Atkinson, K. Cooper, D. A. Ritchie, and A. J. Shields,
Phys. Rev. Lett. \textbf{99}, 266802 (2007);
 A. G. Dijkstra and Y. Tanimura, Phys. Rev. Lett. \textbf{104}, 250401 (2010);
 Z.-h. Tang, G.-x. Li, and Z. Ficek, Phys. Rev A \textbf{82}, 063837 (2010);
 M. Abazari, A. Mortezapour, M. Mahmoudi, and M. Sahrai,
Entropy \textbf{13}, 1541 (2011);
H.-T. Wang, C.-F. Li, Y. Zou, R.-C. Ge, and G.-C. Guo, Physica A \textbf{390}, 3183 (2011); 
D. Boyanovsky, Phys. Rev. A \textbf{87}, 033815 (2013);
S. Bera, S. Florens, H. U. Baranger, N. Roch, A. Nazir, and A. W. Chin,
Phys. Rev. B \textbf{89}, 121108(R) (2014);
 A. Streltsov, U. Singh, H. S. Dhar, M. N. Bera, and G. Adesso
Phys. Rev. Lett. \textbf{115}, 020403 (2015);
 E. Chitambar and M.-H. Hsieh,
Phys. Rev. Lett. \textbf{117}, 020402 (2016); 
 A. Streltsov, E. Chitambar, S. Rana, M. N. Bera, A. Winter, and M. Lewenstein,
Phys. Rev. Lett. \textbf{116}, 240405 (2016);
K. C. Tan and H. Jeong, Phys. Rev. Lett. \textbf{121}, 220401 (2018);
 L.-F. Qiao, J. Gao, A. Streltsov, S. Rana, R.-J. Ren, Z.-Q. Jiao, C.-Q. Hu, X.-Y. Xu, C.-Y. Wang, H. Tang, A.-L. Yang, Z.-H. Ma, M. Lewenstein, and X.-M. Jin, Phys. Rev. A \textbf{98}, 052351 (2018);
S.  Goswami, S. Adhikari, and A. S. Majumdar,  Quantum Inf. Process. \textbf{18}, 36 (2019); 
M. Afrin and T. Qureshi, Eur. Phys. J. D \textbf{73}, 31 (2019);
 Y. Xi, T. Zhang, Z. Zheng, X. Li-Jost, and S.-M. Fei, 
 Phys. Rev. A \textbf{100}  022310 (2019);
L.-H. Ren, M. Gao, J. Ren, Z. D. Wang, and Y.-K. Bai, 
arXiv:2004.03995; 
T. Theurer, S. Satyajit, and M. B. Plenio, 
arXiv:2004.04536.







\end{thebibliography}

\begin{thebibliography}{99}

\bibitem{ja-re-ja-re-uRe-1}
J. {\AA}berg, 
arXiv:quant-ph/0612146 (2006);
%
T. Baumgratz, M. Cramer, and M. B. Plenio, 
Phys. Rev. Lett. \textbf{113}, 140401 (2014);
%
A. Winter and D. Yang, 
Phys. Rev. Lett. \textbf{116}, 120404 (2016);
%
A. Streltsov, G. Adesso, and M. B. Plenio, 
Rev. Mod. Phys. {\bf 89}, 041003 (2017).

\bibitem{boRda-1} 
C. H. Bennett, D. P. DiVincenzo, J. A. Smolin, and W. K. Wootters,
Phys. Rev. A \textbf{54}, 3824 (1996).


\bibitem{neelmadhab} E. M. Rains, quant-ph/9707002; V. Vedral and M. B. Plenio, 1998, Phys. Rev. A \textbf{57}, 1619 (1998).

\bibitem{megh-1} 
R. F. Werner,
Phys. Rev. A \textbf{40}, 4277 (1989).

\end{thebibliography}
\end{document}